# A simple model of mortality trends aiming at universality: Lee Carter + Cohort

Edouard Debonneuil – AXA Cessions, Paris, France

**Summary** The Lee Carter modelling framework is widely used because of its simplicity and robustness despite its inability to model specific cohort effects. A large number of extensions have been proposed that model cohort effects but there is no consensus. It is difficult to simultaneously account for cohort effects and age-adjusted improvements and we here test a simple model that accounts for both: we simply add a non age-adjusted cohort component to the Lee Carter framework. This is a trade-off between accuracy and robustness but, for various countries present in the Human Mortality Database and compared to various models, the model accurately fits past mortality data and gives good backtesting projections.

## Introduction

The modelling and forecasting of mortality trends has received considerable attention since the development of a simple stochastic model by Lee and Carter (1992). This model - that models the log of the force of mortality by an age component plus an age-adjusted period component plus a gaussian error - has been improved in various ways such as by improving the modelling of the error term (Brouhns et al., 2002) or by adding cohort components.

There is no consensus on how to model cohort effects. A good starting point to studying them is the toolkit Lifemetrics (Coughlan et al., 2007), that implements various models that have then been analyzed in details (Cairns et al, 2007, 2008). In particular the model 1 – the one with the Lee Carter parametric structure - is robust but does not account for cohort specificities. The model 2 - that adds an age-adjusted cohort component (Renshaw and Haberman, 2006) - has a tendency towards over fitting and associated lack of robustness. The model 3 – model 2 without age adjustments of period and cohort components (Currie et al., 2006) – is robust but does not account of age-specific mortality improvements.

This article describes a model that is a trade-off between model 2 and model 3, in the aim to be as universal as possible. It models the log of the force of mortality by an age component plus an age-adjusted period component plus a non age-adjusted cohort effect: it can be viewed as a Lee Carter model plus a simple cohort component or as a constrained Renshaw Haberman model. One application of such a model could be to fit many cases: it adapts to some age-specific improvements and to some cohort effects, but due to its limited number of parameters it should limit the risk of over fitting.

To empirically test this degree of universality we will test it on mortality trends of various countries, using historical mortality data from general populations (Human Mortality Database).

## The LCC model: Lee Carter + Cohort

### Definition

Using the notations of lifemetrics, figure 1 describes the new model (that we will name M9, or LCC) along with the models already presented in Lifemetrics. To be uniquely defined, constraints must be added: similarly to the models described in Lifemetrics we will require $\beta_x^{(2)}$ to sum to 1, $\kappa_t^{(2)}$ to sum to 0 and $\gamma_{t-x}^{(3)}$ to sum to 0. It was implemented as a straightforward derivation of the implementation for the M2 model: $\beta_x^{(3)}$ is simply set to 1 and not fitted.

| Model | formula |
|---|---|
| M1 | $\log m(t,x) = \beta_x^{(1)} + \beta_x^{(2)}\kappa_t^{(2)}$ |
| M2 | $\log m(t,x) = \beta_x^{(1)} + \beta_x^{(2)}\kappa_t^{(2)} + \beta_x^{(3)}\gamma_{t-x}^{(3)}$ |
| M3 | $\log m(t,x) = \beta_x^{(1)} + n_a^{-1}\kappa_t^{(2)} + n_a^{-1}\gamma_{t-x}^{(3)}$ |
| M4 | $\log m(t,x) = \sum_{i,j} \theta_{ij} B_{ij}^{ay}(x,t)$ |
| M5 | $\text{logit } q(t,x) = \kappa_t^{(1)} + \kappa_t^{(2)}(x-\bar{x})$ |
| M6 | $\text{logit } q(t,x) = \kappa_t^{(1)} + \kappa_t^{(2)}(x-\bar{x}) + \gamma_{t-x}^{(3)}$ |
| M7 | $\text{logit } q(t,x) = \kappa_t^{(1)} + \kappa_t^{(2)}(x-\bar{x}) + \kappa_t^{(3)}\left((x-\bar{x})^2 - \hat{\sigma}_x^2\right) + \gamma_{t-x}^{(4)}$ |
| M8 | $\text{logit } q(t,x) = \kappa_t^{(1)} + \kappa_t^{(2)}(x-\bar{x}) + \gamma_{t-x}^{(3)}(x_c - x)$ |
| M9 | $\log m(t,x) = \beta_x^{(1)} + \beta_x^{(2)}\kappa_t^{(2)} + \gamma_{t-x}^{(3)}$ |

Figure 1. Lifemetrics models and the Lee Carter+Cohort model (M9)

**Fitting**

The parameters β κ γ are fitted by maximum likelihood. The Bayesian Information Criterion (BIC) metric can be used to estimate whether a model fits data well: that indicator takes into account the fact that a model with many parameters has more chances to fit a set of data by chance. Table 1 provides the obtained BIC when fitting historical deaths and exposures of various countries for males and females aged 40 to 95 year old between 1955 and 1980. M9 fits well as it leads to high BIC:

| Country, gender | M1 | M2 | M3 | M5 | M6 | M7 | M8 | M9 |
|---|---|---|---|---|---|---|---|---|
| UK Male | -10714 | -9182 | -9215 | -24645 | -9808 | -9202 | -9266 | **-9014** |
| UK Female | -9574 | -9198 | -9450 | -19336 | -11182 | -9788 | -9846 | **-9069** |
| US Male | -13844 | **-11491** | -12951 | -27286 | -19093 | -14818 | -15964 | -11602 |
| US Female | -14115 | **-11163** | -12648 | -65996 | -16634 | -16547 | -16032 | -11415 |
| France Male | -10811 | -9080 | -8992 | -14256 | -10148 | -9166 | -9538 | **-8943** |
| France Female | -9147 | -8992 | -9010 | -42131 | -13648 | -11080 | -10761 | **-8846** |
| Switz. Male | -6773 | -7092 | -6806 | -6541 | **-6715** | -6775 | -6700 | -6935 |
| Switz. Female | **-6743** | -7060 | -6772 | -8230 | -6977 | -6877 | -6774 | -6923 |
| Spain Male | -9058 | -8538 | -9145 | -11237 | -10595 | -10183 | -10041 | **-8509** |
| Spain Female | -9103 | -8561 | -9809 | -22162 | -15948 | -12517 | -13131 | **-8533** |
| Japan Male | -11465 | -9249 | -9274 | -17559 | -9656 | **-8851** | -9242 | -9096 |
| Japan Female | -11374 | -9194 | -9536 | -38082 | -14092 | -10754 | -10750 | **-9140** |

Table 1: on average M9 has a better BIC fitting indicator than M1, M2, M3, M5, M6, M7 and M8. For each country the best BIC value is set in bold.

## The UK as an example

Here are a few examples to visually describe some limits of the models:

- Cohort effect: as shown by the $\gamma_{t-x}^{(3)}$ curves in figure 2a there is clearly a strong cohort effect in the UK for people born around 1920 (the curves for women are very similar; data not shown). The Lee Carter framework (M1) does not account for it; this is probably why M1 instead provides a $\beta_x^{(2)}$ that is very surprising, much less likely than the $\beta_x^{(2)}$ provided by M2 or M9, as shown in figure 2b.

- Age-adjusted mortality improvements: the latter $\beta_x^{(2)}$ provided by M2 and M9 suggest that improvements mostly affect older ages; M3 cannot account for it because $\beta_x^{(2)}=1$ for M3

- Overfitting in the more complex model M2: No obvious overfitting was found for the UK for dates between 1955 and 1980 but some is found for dates between 1981 and 2006 for females, as shown in blue in figure 2c: the very perturbed $\beta_x^{(2)}$ and $\beta_x^{(3)}$ probably compensate each other to fit the data as much as possible, but those shapes are clearly artifacts (they would suggests much higher mortality improvements for age 47 than for age 46).

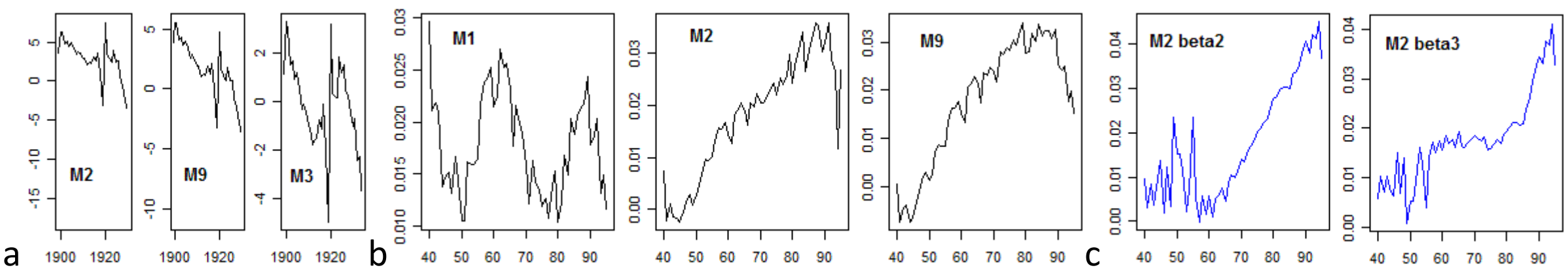

Figure 2: a) cohort effect ($\gamma_{t-x}^{(3)}$) fitted by M2, M3 and M9 for males: UK has a clear cohort effect. b) age-distribution of mortality improvements ($\beta_x^{(2)}$) provided by M1, M2 and M9 for males: M1 provides improbable shapes. c) $\beta_x^{(2)}$ and $\beta_x^{(3)}$ provided by M2 in the case of females aged 40 to 95 and between 1981 and 2006: there is a clear overfitting.

## Projections of M1, M2, M3 and M9

The projection of models M1, M2, M3 and M9 requires extrapolating $\kappa_t^{(2)}$ after 1980. There are many manners to do so. Rather than choosing a specific manner we computed what $\kappa_t^{(2)}$ should be, to match mortality rates between 1981 and 2006 as much as possible: the parameters that were previously fitted ($\beta_x^{(1)}$, $\beta_x^{(2)}$, $\beta_x^{(3)}$ and $\gamma_{t-x}^{(3)}$ for each model, country and gender between 1955 and 1980) are reused between 1981 and 2006 except $\kappa_t^{(2)}$ that is then fitted by maximum likelihood to match mortality rates between 1981 and 2006 (restricted to cohorts that were present in the data between 1955 and 1980). Figure 3 shows the corresponding $\kappa_t^{(2)}$ along with linear extrapolation of past $\kappa_t^{(2)}$ (norm 1 in red, norm 2 in green, norm infinite in grey).

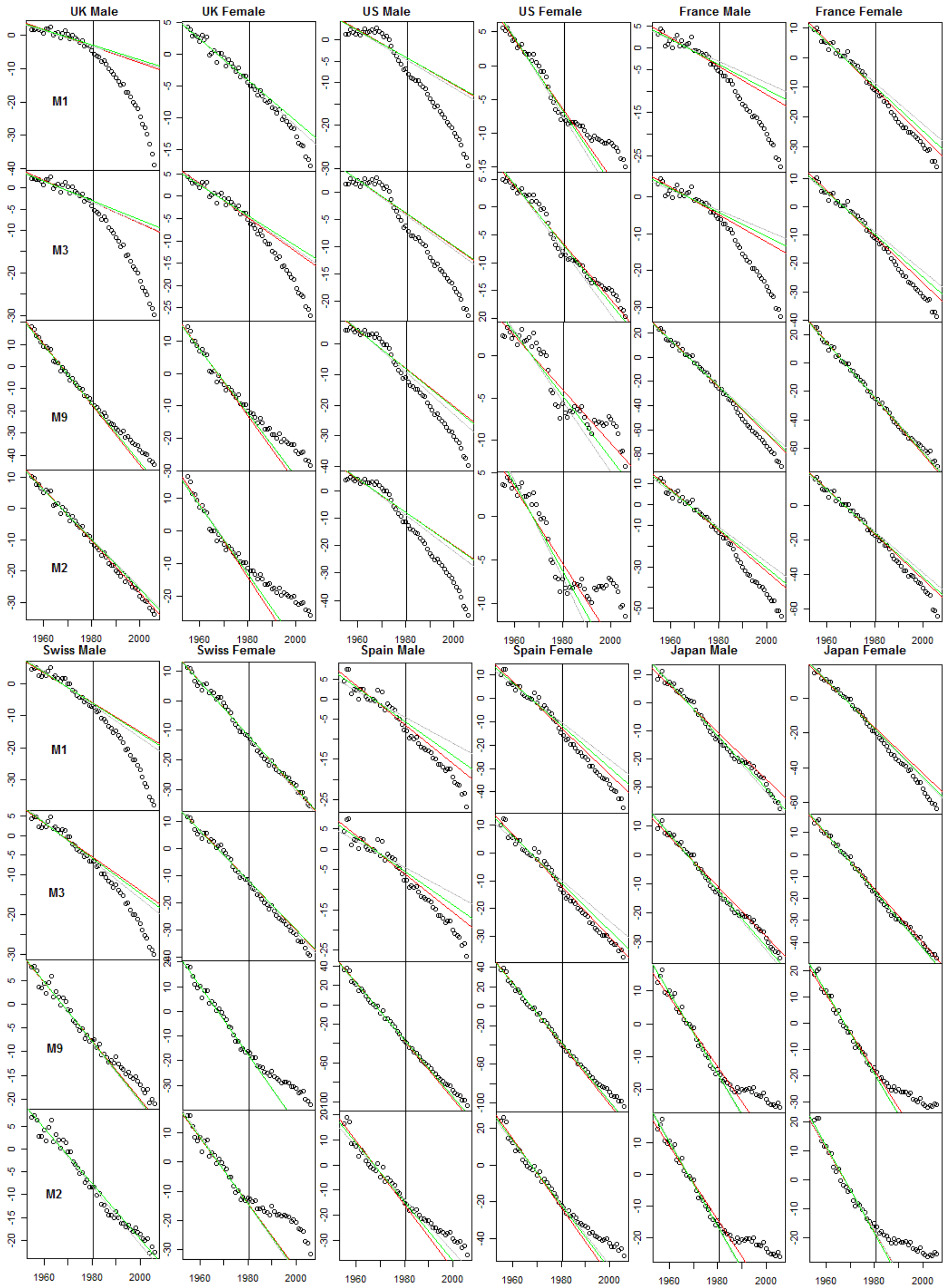

Figure 3. Fitted $\kappa_t^{(2)}$ and goodness of projection.

**Discussion and conclusion**

Not all models predicted correctly all cases. All countries and times are specific. The UK is particular for its 1920-born short-lived cohort. Switzerland data is particular for its young-age-only mortality improvements, etc. Model M9 seems to fit mortality data quite well: it can be viewed as a trade-off between accuracy and robustness. Concerning projections, linear projections of $\kappa_t^{(2)}$ do not seem to be the best solution in all cases but it seems difficult to have a better choice without referring to some expert judgement. Model M9 is arguably better in terms of projections. For a given country it shows similar evolutions of $\kappa_t^{(2)}$ for males and females, which suggests that it captures the mortality trends of countries, even if those trends are difficult to model in a generic manner.

It is difficult to expect much better as the discussed models do not take into account the background such as biomedical improvements and prevention knowledge that lead to longer lives. For example the slowdown of mortality improvements in Japan could be a reality due to a partial westernization of the Japanese society, and this westernization cannot be predicted by blind mathematical models.

It is therefore in the absence of such contextual modelling that we suggest the choice of the Lee Carter + Cohort model for "acontextual" mortality trend modelling.